\documentclass{optica-article}

\journal{opticajournal} 

\articletype{Research Article}

\usepackage{lineno}
\usepackage{subfigure}
\usepackage{siunitx}
\usepackage{gensymb}
\usepackage{tabularx}

\begin{document}

\title{Two-colour balanced optical cross-correlator using fibre-coupled PPLN waveguides}

\author{Jonathan Christie,\authormark{1,2,*} James R. Henderson,\authormark{2,\textdagger} Edward W. Snedden,\authormark{2} and Laura Corner\authormark{1}}

\address{\authormark{1}Department of Materials, Design and Manufacturing Engineering, University of Liverpool, Brownlow Hill, Liverpool L69 3GH, UK\\
\authormark{2}Accelerator Science and Technology Centre, Science and Technology Facilities Council, Daresbury Laboratory, Sci-Tech Daresbury, Keckwick Lane, Warrington WA4 4AD, UK\\
\authormark{\textdagger}Current affiliation: Coherent Scotland Ltd, West of Scotland Science Park, Maryhill Rd, Glasgow G20 0XA, UK}

\email{\authormark{*}j.s.christie@liverpool.ac.uk} 


\begin{abstract*} 

\noindent We present a two-colour fully fibre-coupled balanced optical cross-correlator (BOXC) based on sum-frequency generation (SFG) between \qty{1560}{nm} and \qty{800}{nm} laser pulses using waveguides implemented in type-0 phase-matched periodically poled LiNbO$_{3}$ (PPLN) crystals. The interaction has an effective nonlinear coefficient of $d_{\text{eff}}$ = \qty{16.1}{pm/V}, many times higher than comparable nonlinear crystals used for this SFG interaction such as barium borate (BBO). The resulting sensitivity of the cross-correlator is measured to be \qty{5.11}{mV/fs}, five times greater than current bulk-optic BOXCs after accounting for differences in transimpedance gain and photodetector responsivity, with the potential for significantly higher sensitivity after optimisations to the cross-correlator design.

\end{abstract*}

\section{Introduction} \label{sec:intro}

In modern accelerator and free-electron laser (FEL) facilities, ultra-precise timing distribution systems play a crucial role. Many accelerator-laser experiments, such as pump-probe spectroscopy and laser wakefield acceleration, require few-femtosecond synchronisation between the accelerated electron bunches and the experiment laser \cite{xfel-use-3, Reitsma2005}, which cannot be achieved using electrical synchronisation methods \cite{Glownia2010}. To achieve the necessary synchronisation performance, several FELs and accelerators use a low-noise mode-locked laser, known as the optical master oscillator (OMO), as a system-wide clock. Pulses from the OMO are distributed throughout the accelerator facility via stabilised fibre links which counteract optical path length fluctuations due to changes in temperature and humidity. This allows for the distribution of optical pulses over kilometre-long accelerator facilities with few-femtosecond arrival time jitter \cite{xin2018, schulz:fel2019-web04}, which can then be used to trigger the necessary accelerator and laser subsystems.

To take advantage of the stability of the OMO pulse train for accelerator-laser experiments, a balanced optical cross-correlator (BOXC) is typically used to measure the relative timing jitter between the pulse trains from the experiment laser and the OMO. The BOXC uses a nonlinear crystal to determine the relative time separation between two laser pulses from the intensity of the generated sum-frequency radiation, and the resulting error signal from the BOXC can be used to synchronise laser sources to less than \qty{10}{fs} \cite{schulz2015, clli:ibic21-wepp05}. However, many current BOXCs utilise bulk free-space nonlinear crystals, requiring free-space optics to direct and focus the laser pulses within the crystal. This makes them susceptible to misalignment of the free-space optics over time due to environmental fluctuations, thus limiting their long-term stability. With modern X-ray FELs (XFELs) being capable of producing attosecond X-ray pulses \cite{Kang2020, Duris_2019} and growing interest in particle physics on attosecond time scales \cite{attosecond-physics}, future accelerator facilities will require a laser-to-laser synchronisation system that is more sensitive to timing jitter and more resilient to environmental changes.

These issues can be addressed by a fully fibre-coupled BOXC, where the laser pulses travel through optical fibres and fibre-coupled components to interact in a fibre-coupled nonlinear crystal. This significantly reduces the number of free-space components, as demonstrated by existing one-colour fully fibre-coupled BOXCs \cite{Callahan2014, Safak:22}, potentially making the system more robust against environmental fluctuations and less challenging to keep aligned over long time scales. The one-colour fully fibre-coupled BOXC uses a nonlinear crystal with an integrated waveguide, greatly increasing the conversion efficiency of the crystal compared to bulk crystals \cite{nejadmalayeri2009}. This allows them to generate significantly more sum-frequency radiation, with the resulting BOXC being up to 100 times more sensitive to timing jitter than comparable bulk-optic BOXCs \cite{Callahan2014, Safak:22}. However, a two-colour fully fibre-coupled BOXC has not yet been demonstrated to the best of our knowledge. With the sensitivity performance that has been achieved by one-colour fully fibre-coupled BOXCs, a two-colour fully fibre-coupled BOXC could allow for sub-femtosecond laser-to-laser synchronisation between lasers of different wavelengths, such as the OMO and high-power experiment lasers at many accelerator and XFEL facilities \cite{lcls-II, euxfel, clara-specs}. 

In this paper, we present a two-colour fully fibre-coupled BOXC based on sum-frequency generation (SFG) between laser pulses from the \qty{1560}{nm} OMO and an \qty{800}{nm} experiment laser. The BOXC uses two \qty{5}{mm}-long, type-0 phase-matched, periodically poled LiNbO$_{3}$ (PPLN) crystal waveguides from HC Photonics to generate the BOXC error signal. After simulating the performance of the BOXC using SNLO \cite{snlo}, we measure the conversion efficiency of the two PPLN waveguides and the sensitivity of the BOXC to timing fluctuations. The current BOXC is found to be several times more sensitive than comparable bulk-optic BOXCs, with the possibility of optimisations to the sensitivity in the future.


\section{Principle of operation and theory} \label{sec:principle-of-operation}

\begin{figure}[b!]
    \centering
    \includegraphics[width=\textwidth]{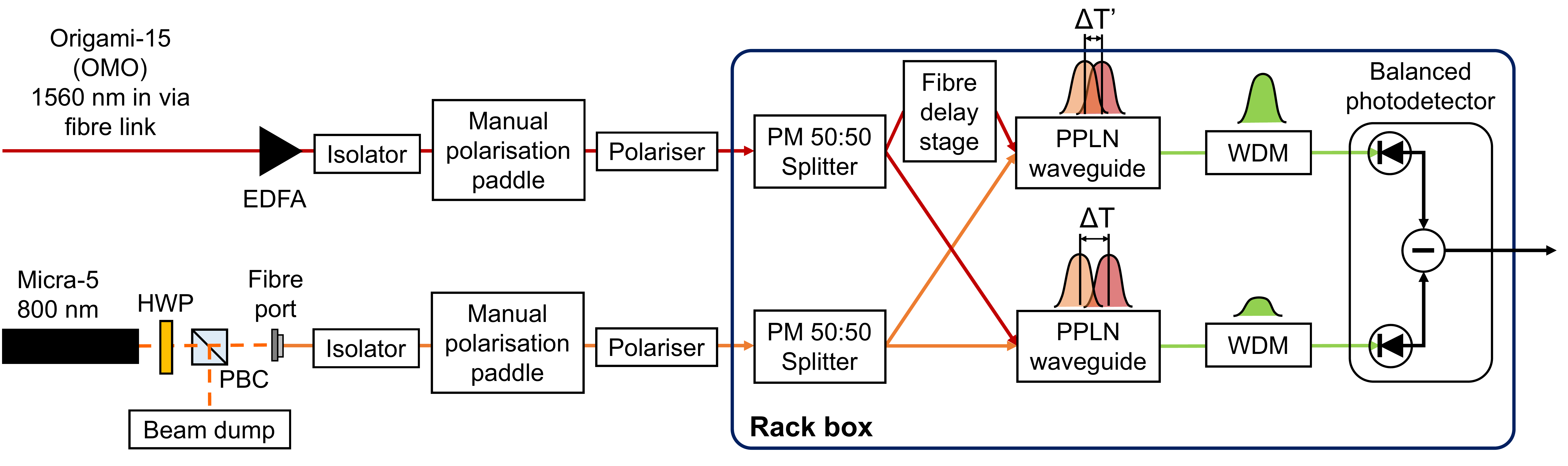}

    \caption{Layout of the two-colour fully fibre-coupled BOXC. The fibre delay stage changes the optical path length of one of the \qty{1560}{nm} pulse components (red), resulting in the pairs of \qty{1560}{nm} and \qty{800}{nm} (orange) pulses having different temporal overlaps at the entrance of the two PPLN waveguides. The generated sum-frequency pulses (green) will then have different intensities, and the difference in the voltage signal from the sum-frequency pulses is proportional to the time separation of the input pulses before the splitters. EDFA: Erbium-doped fibre amplifier; HWP: Half-wave plate; PBC: Polarising beamsplitter cube; WDM: Wavelength division multiplexer; Rack box: \qty{48.3}{cm} $\times$ \qty{48.3}{cm} rack box housing a \qty{39.4}{cm} $\times$ \qty{40.6}{cm} optical breadboard.}
    \label{fig:tc-fibre-boxc-layout}
\end{figure}

The principle of operation of the two-colour fully fibre-coupled BOXC design is shown in Figure \ref{fig:tc-fibre-boxc-layout}. \qty{1560}{nm} pulses from the Origami-15 OMO, which operates at a repetition rate of $f_{1560}$ = \qty{249.875}{MHz}, travel to the BOXC enclosure via a stabilised fibre link. The pulses from the link, which have a full width half maximum (FWHM) spectral width of \qty{10.9}{nm} and a FWHM pulse duration of \qty{250}{fs}, are amplified by an erbium-doped fibre amplifier (EDFA). The Micra-5, which operates at a repetition rate of $f_{800} = f_{1560}/3$ = \qty{83.292}{MHz}, produces \qty{800}{nm} pulses with a FWHM spectral width of \qty{27.4}{nm} and a measured FWHM pulse duration of \qty{970}{fs}, much greater the sech$^{2}$ bandwidth-limited duration of \qty{24.5}{fs}; this is due to both misalignment of the dispersion-compensating prisms within the Micra-5 oscillator and dispersive optical elements within the beam transport to the BOXC enclosure. These pulses pass through a half-wave plate and polarising beamsplitter cube to adjust their power before being coupled into single-mode optical fibre. Both wavelength pulses travel to separate fibre polarisation paddles and polarisers to ensure they have the correct polarisation for the PPLN waveguides. Polarisation-maintaining (PM) fibre is used after the paddles to prevent polarisation changes within the optical fibre. The pulses then travel to a \qty{48.3}{cm} $\times$ \qty{48.3}{cm} rack box housing the electrical devices, where they are split by fibre-coupled 50:50 splitters before travelling to separate PPLN waveguides. One \qty{1560}{nm} pulse component travels to a fibre delay stage, whereas the other component travels through a fixed length of fibre such that, when the delay stage is in its centre position, both \qty{1560}{nm} pulse components travel the same optical path length. Thus, when the delay stage is moved away from its centre position, it imparts a relative time delay $D$ between the \qty{1560}{nm} pulse components. This results in the sets of \qty{800}{nm} and \qty{1560}{nm} pulses having different pulse delays $\Delta T$ and $\Delta T' = \Delta T + D$, where the pulse delay is the time between the intensity peaks of the \qty{800}{nm} and \qty{1560}{nm} pulses at the entrance of the PPLN waveguide (note that the pulse delay is defined to be positive when the \qty{1560}{nm} pulse is trailing the \qty{800}{nm} pulse). The two waveguides then generate separate \qty{528.8}{nm} sum-frequency pulses with different intensities due to the different temporal overlaps of the input pulses. After passing through a wavelength division multiplexer (WDM) to remove any background \qty{800}{nm} and \qty{1560}{nm} radiation passing through the waveguide, the sum-frequency pulses are measured by a balanced photodetector and the difference in the two voltage responses is taken. Scanning over $\Delta T$ gives a voltage error signal that is linearly proportional to the time separation of the pulses about the zero-crossing point of the error signal. This signal can be used to adjust the cavity length and hence repetition rate of the Micra-5, with the aim of keeping the error signal locked to the zero-crossing point, resulting in the synchronisation of the Micra-5 to the OMO pulse train.

\begin{figure}[b!]
    \centering
    \subfigure[]{\includegraphics[width=0.495\textwidth]{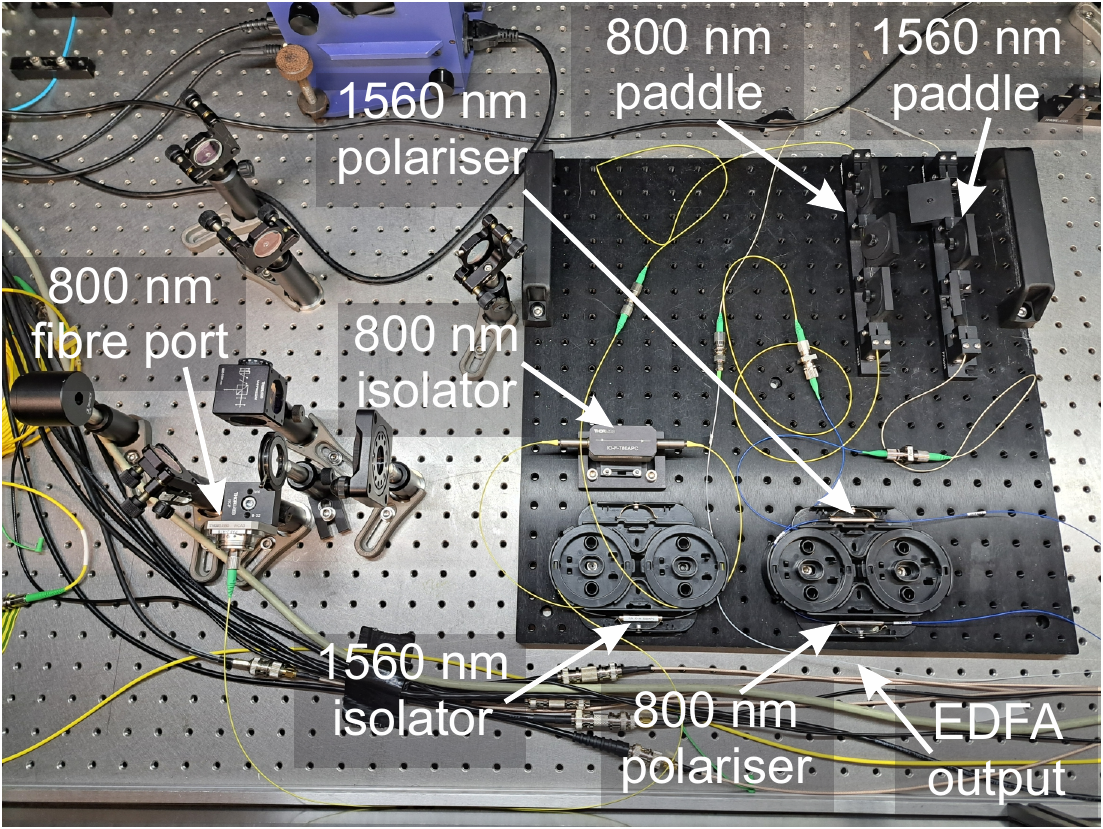}}
    \subfigure[]{\includegraphics[width=0.495\textwidth]{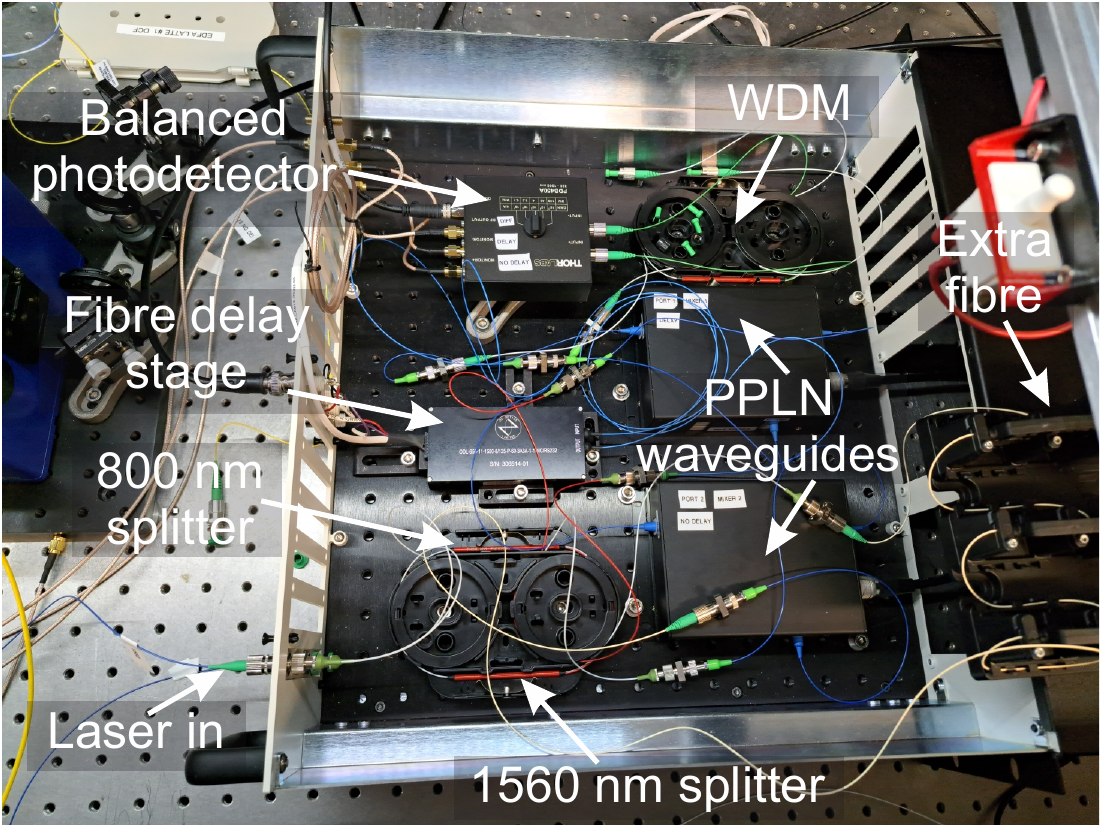}}
    \caption{Images of the two-colour fully fibre-coupled BOXC. (a) The polarisation control section of the BOXC, which is housed on a \qty{45.0}{cm} $\times$ \qty{45.0}{cm} optical breadboard and contains the isolators, polarisation paddles, and inline polarisers. (b) The cross-correlation section of the BOXC, which is housed on a \qty{39.4}{cm} $\times$ \qty{40.6}{cm} optical breadboard inside a rack box and contains the 50:50 splitters, fibre delay stage, PPLN waveguides, and balanced photodetector. WDM: Wavelength division multiplexer.}
    \label{fig:lab-layout}
\end{figure}

Figure \ref{fig:lab-layout} shows the BOXC layout in the laboratory. The BOXC components are housed on a \qty{45.0}{cm} $\times$ \qty{45.0}{cm} optical breadboard and on a \qty{39.4}{cm} $\times$ \qty{40.6}{cm} optical breadboard inside a rack box; future BOXC designs will aim to house all of the fibre-coupled components in a single rack box to minimise the footprint of the BOXC. A motorised polarisation paddle is used to match the optical path length of the \qty{1560}{nm} fibre delay stage in its centre position; this will be replaced by a fixed length of PM fibre in future designs. 

Each PPLN waveguide module consists of a \qty{5}{mm}-long PPLN crystal with an integrated ridge waveguide, components to couple the \qty{1560}{nm} and \qty{800}{nm} pulses from the input fibres into the waveguide and to couple the sum-frequency pulses from the waveguide into the output fibre, and a thermoelectric cooler (TEC) that can vary the waveguide temperature between the set limits of \qty{15.0}{\degreeCelsius} and \qty{70.0}{\degreeCelsius}. The waveguide modules have separate input and output ports using \qty{0.3}{m}-long PM fibre pigtails matched to each of the three wavelengths in the nonlinear interaction, ensuring that losses due to mode mismatch are minimised. Additionally, the crystals are anti-reflection coated to further reduce losses. To guide the laser pulses over the length of the PPLN crystal, a ridge waveguide is implemented such that the PPLN ridge is surrounded by low-index cladding \cite{Paschottachannel_waveguides}. This confines the pulses to the ridge waveguide and prevents them from diffracting within the crystal, allowing for longer interaction lengths, greater laser intensity over the length of the crystal, and thus increased sum-frequency conversion efficiency compared to bulk-optic crystals. PPLN was chosen for its high transparency at the interaction wavelengths and for its large effective nonlinear coefficient for type-0 phase-matching of $d_{\text{eff}}$ = \qty{16.1}{pm/V} \cite{handbook-nonlinear-crystals, ppln-coefficient}. This is around 5 times greater than that of bulk barium borate (BBO) \cite{handbook-nonlinear-crystals}, which is typically used for bulk-optic two-colour BOXCs \cite{schulz2015, clli:ibic21-wepp05}.


\subsection{Fibre dispersion} \label{subsec:fibre-dispersion}

In this current version of the two-colour fully fibre-coupled BOXC design, the \qty{800}{nm} pulses travel through \qty{8.7}{m} of single-mode fibre. The \qty{1560}{nm} pulses travel through approximately \qty{12}{m} of single-mode fibre and \qty{3.0}{m} of erbium-doped fibre from the EDFA, as well as \qty{9.1}{m} of single-mode fibre from the BOXC components, giving a total fibre length of approximately \qty{24.1}{m}. Due to their large spectral widths, the pulses undergo significant temporal broadening as a result of fibre dispersion. For the \qty{800}{nm} pulses, the measured pulse duration before the waveguide input fibre is approximately \qty{39}{ps}, much longer than the initial pulse duration of \qty{970}{fs}. Likewise for the \qty{1560}{nm} pulse, the pulse duration after the EDFA and BOXC is measured to be \qty{1.1}{ps}, compared to the initial \qty{250}{fs} duration from the link. To quantify the amount of dispersion experienced by the pulses, the group delay dispersion and linear chirp need to be determined; a full derivation of the following expressions can be found in \cite{yariv1997optical}.

Consider an initially bandwidth-limited Gaussian laser pulse with centre frequency $\omega_{0}$, centre wavelength $\lambda$, and FWHM pulse duration $\tau_{0}$ travelling through an optical fibre of length $L$. The expression for the \textit{group velocity dispersion} (GVD) of the fibre is

\begin{equation} \label{eq:gvd}
    \text{GVD} = \beta_{2} = - \frac{\lambda^{2}}{2 \pi c_{0}} D(\lambda) \; ,
\end{equation}

\noindent where $c_{0}$ is the speed of light in vacuum and $D(\lambda)$ is the \textit{dispersion parameter} of the fibre. The \textit{group delay dispersion} (GDD) is thus given by

\begin{equation} \label{eq:gdd}
    \text{GDD} = \beta_{2} L \;.
\end{equation}

\noindent The dispersion of the optical fibre causes the pulse to broaden temporally as it propagates through the fibre, with the broadened FWHM pulse duration $\tau(L)$ given by

\begin{equation} \label{eq:dispersive-pulse-broadening}
    \tau(L) = \tau_{0} \sqrt{1+\bigg(\frac{2 \ln 2}{\pi c_{0}} \frac{DL\lambda^{2}}{\tau_{0}^2} \bigg)^{2}} = \tau_{0} \sqrt{1+\bigg(4 \ln 2 \frac{\beta_{2} L}{\tau_{0}^2} \bigg)^{2}} \;.
\end{equation}

In addition to increasing the pulse duration, the dispersion of the optical fibre will also introduce a linear chirp to the pulse, where the instantaneous frequency of the pulse changes linearly with time. The magnitude of the linear chirp, $b$, is given by the expression

\begin{equation} \label{eq:linear-chirp-parameter}
    b = \frac{4 \beta_{2} z}{2 \pi (1/\alpha_{\tau}^{2} + 4 \beta_{2}^{2} z^{2})} \; , 
\end{equation}

\noindent where $\alpha_{\tau} = 2 \ln 2 / \tau_{0}^{2}$. 

Substituting the \qty{800}{nm} optical fibre dispersion of \qty{-118}{ps/nm*km} \cite{800nm-fibre-dispersion} into Equation \ref{eq:gvd} gives a GVD of \qty{4.0E4}{fs^{2}/m} and a GDD of \qty{3.5E5}{fs^{2}} over the total fibre length of \qty{8.7}{m}. After accounting for the initial pulse not being bandwidth limited, the expected pulse duration at the PPLN waveguide is around \qty{39.6}{ps} with a positive chirp of \qty{0.45}{THz/ps}, close to the measured value of \qty{39}{ps}. 

Similarly for the \qty{1560}{nm} pulses, using a fibre dispersion of \qty{18}{ps/nm*km} \cite{1560nm-smf-28-fibre-dispersion} and total fibre length of \qty{24.1}{m}, the GVD and GDD are \qty{-2.3E4}{fs^{2}/m} and \qty{-5.6E5}{fs^{2}} respectively. Thus, the expected pulse duration is \qty{6.2}{ps} with a negative chirp of \qty{-0.28}{THz/ps} at the PPLN waveguide, different from the measured value of \qty{1.1}{ps}. This difference is likely due to a number of factors, such as self-phase modulation and other nonlinear effects reducing the rate of dispersion \cite{Zheltikov:18} and not knowing the dispersion of the erbium-doped fibre used in the EDFA; investigating the source of this difference will be a focus of future work.


\section{Effect of chirp on sum-frequency generation} \label{sec:chirp-sfg}

Despite both the \qty{800}{nm} and \qty{1560}{nm} pulses becoming longer in time due to fibre dispersion, the chirps of the two pulses are opposite in direction to each other. Thus, instead of the sum-frequency interaction occurring over the entire pulse duration, as is the case for unchirped pulses, the sum-frequency interaction is instead limited to a small portion of the pulse bandwidth where the phase-matching condition of the nonlinear crystal is satisfied \cite{Raoult:98}. This reduces the range of pulse delays where the sum-frequency interaction strongly occurs, allowing for the generation of cross-correlation traces much narrower than the input pulse durations. 

For a general sum-frequency interaction between two linearly chirped pulses, assuming self-phase modulation effects are negligible for both pulses, the instantaneous frequencies of the two pulses $\omega_{1,2}(t)$ can be written as

\begin{equation} \label{eq:instant-freq-1560nm}
    \omega_{1}(t) = \omega_{10} + b_{1}(t - \tau) \;,
\end{equation}

\begin{equation} \label{eq:instant-freq-800nm}
    \omega_{2}(t) = \omega_{20} + b_{2}t \;,
\end{equation}

\noindent where $\omega_{10}$ and $\omega_{20}$ are the centre frequencies of the unchirped pulses, $b_{1,2}$ are their respective linear chirps, and $\tau$ is the temporal offset between the two pulses. The instantaneous frequency of the sum-frequency pulse $\omega_{3}(t)$ is given by the sum of the two instantaneous frequencies,

\begin{equation} \label{eq:instant-freq-sfg}
    \omega_{3}(t) = \omega_{1}(t) + \omega_{2}(t) = \omega_{30} + (b_{1} + b_{2})t - b_{1} \tau \;,
\end{equation}

\noindent where $\omega_{30} = \omega_{10} + \omega_{20}$. Thus, the sum-frequency pulse will also have a time-dependent instantaneous frequency that depends on the chirp of the two input pulses, $b_{3} = b_{1} + b_{2}$. Due to the phase-matching condition of the nonlinear crystal, which requires the frequency shift $\Delta \omega_{3} = \omega_{3}(t) - \omega_{30} \approx 0$, the chirps of the input pulses can be used to alter the instantaneous frequency of the sum-frequency pulse so that the phase-matching condition is no longer satisfied, effectively gating the sum-frequency process.


\subsection{Simulations of BOXC performance} \label{subsec:boxc-simulation}

To investigate how the \qty{800}{nm} and \qty{1560}{nm} pulse chirps affect the performance of the two-colour fully fibre-coupled BOXC, the `Plane-wave short-pulse mixing' module in SNLO is used to simulate the cross-correlation of the \qty{1560}{nm} and \qty{800}{nm} pulses by one of the PPLN crystals used in the BOXC. 

\begin{figure}[t]
    \centering
    \includegraphics[width=\textwidth]{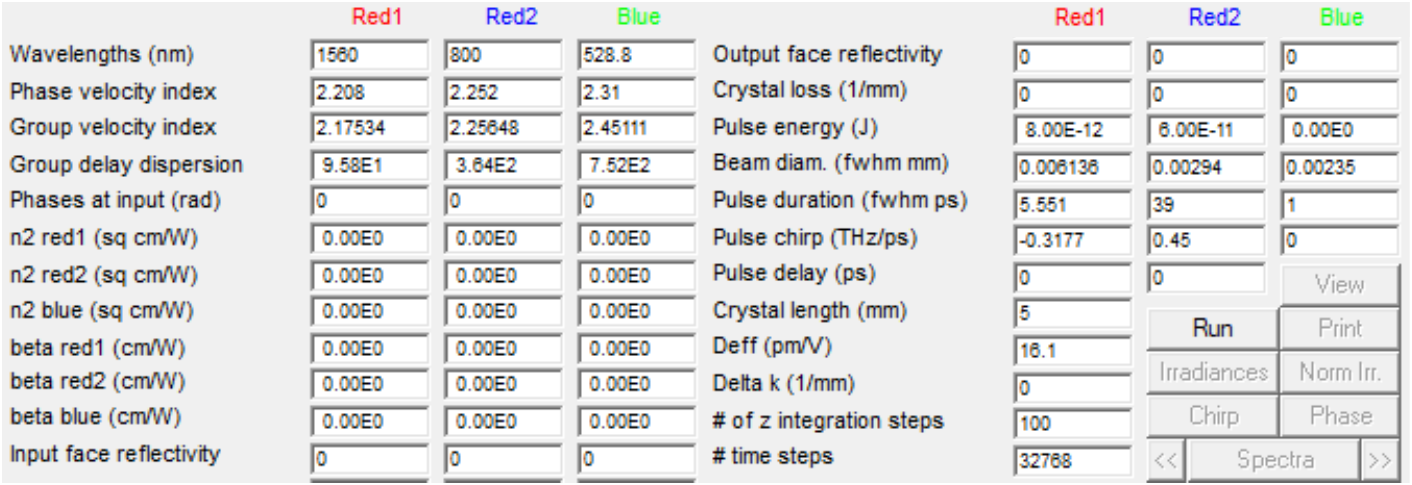}
    \caption{SNLO parameters used to simulate the cross-correlation of the \qty{1560}{nm} and \qty{800}{nm} pulses by one of the PPLN crystals used in the BOXC.}
    \label{fig:snlo-parameters}
\end{figure}

A summary of the crystal and laser parameters used is given in Figure \ref{fig:snlo-parameters}. The GDD of the \qty{800}{nm} pulse is kept constant at \qty{3.5E5}{fs^{2}} to match the measured value, giving a pulse duration and chirp of \qty{39}{ps} and $b_{2}$ = \qty{0.45}{THz/ps} respectively. To simulate the effect of varying the pulse length by adding additional dispersion-compensating fibre and single-mode fibre, the GDD of the \qty{1560}{nm} pulse is changed from \qty{-10E5}{fs^{2}} to \qty{5.0E5}{fs^{2}} and, using a bandwidth-limited duration of \qty{250}{fs}, the corresponding pulse duration and pulse chirp values are calculated using Equations \ref{eq:dispersive-pulse-broadening} and \ref{eq:linear-chirp-parameter} (note that the `Group delay dispersion' row in Figure \ref{fig:snlo-parameters} refers to the GDD \textit{imparted by the crystal}, not the GDD of the pulse entering the crystal). The phase velocity index, group velocity index, and group delay dispersion of the crystal are calculated using the `Quasi-phasematch' module. The nonlinear refractive indices, absorption coefficients, reflectivities, and crystal losses are chosen to be 0 to simplify the simulations. The average powers of the \qty{1560}{nm} and \qty{800}{nm} pulses are chosen to be \qty{2}{mW} and \qty{5}{mW} respectively to be similar to the measured values entering the waveguide connected to the fibre delay stage (see Section \ref{subsec:conversion-efficiency}). This corresponds to respective pulse energies of \qty{8.0}{pJ} and \qty{60}{pJ}. The beam diameters are calculated from the mode-field diameters of the three fibres entering and exiting each PPLN waveguide module, and the effective nonlinear coefficient and crystal length of \qty{16.1}{pm/V} and \qty{5.0}{mm} respectively are provided by the crystal manufacturer.

\begin{figure}[t]
    \centering
    \includegraphics[width=\textwidth]{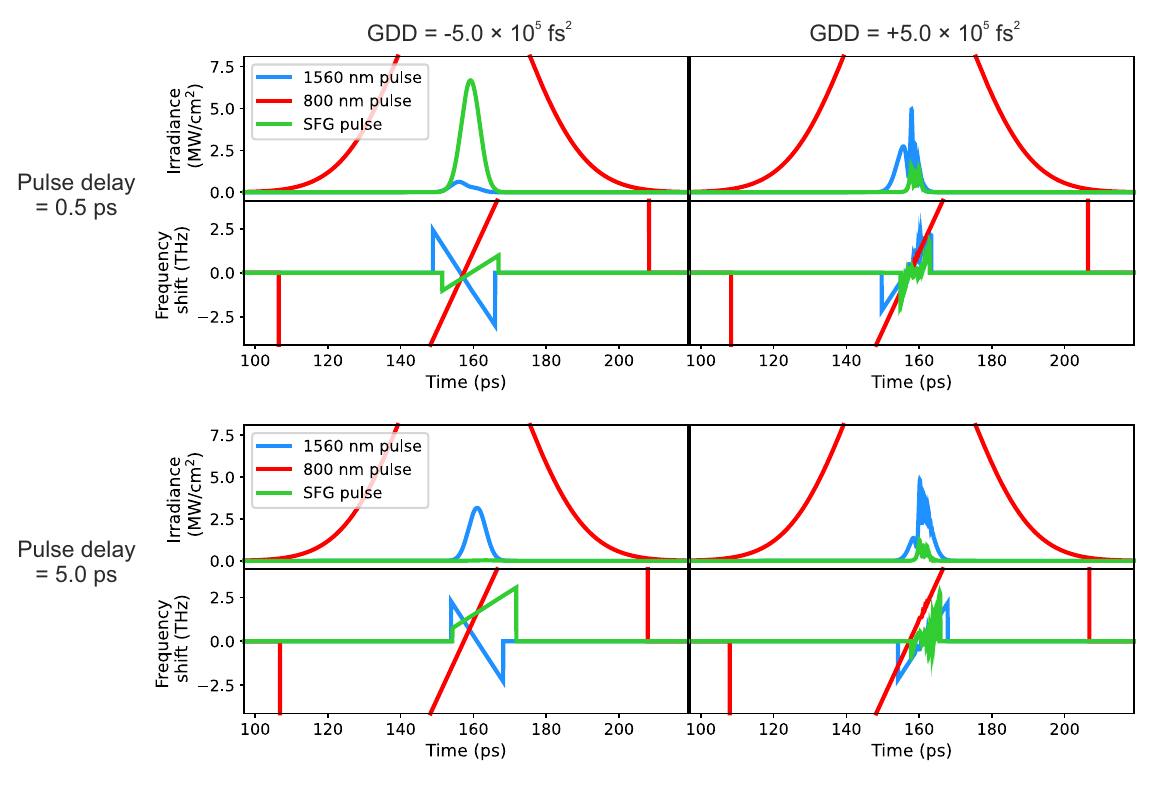}
    \caption{Comparison of the irradiances and frequency shifts of the \qty{1560}{nm}, \qty{800}{nm}, and the \qty{528.8}{nm} (`SFG pulse') sum-frequency pulses at different pulse delays and \qty{1560}{nm} pulse group delay dispersions. Additional pulse parameters are given in Figure \ref{fig:snlo-parameters}.}
    \label{fig:snlo-sfg}
\end{figure}

Figure \ref{fig:snlo-sfg} illustrates how the chirp of the input pulses affects the irradiance and frequency shift of the sum-frequency pulse using the parameters in Figure \ref{fig:snlo-parameters}. In these simulations, the \qty{1560}{nm} pulses have the same GDD magnitude of \qty{5.0E5}{fs^{2}}, corresponding to a chirp magnitude of \qty{0.32}{THz/ps}, but opposite chirp directions. Note that, due to the much greater peak irradiance and pulse duration of the \qty{800}{nm} pulse, the plots are zoomed in so that the irradiances and frequency shifts of the \qty{1560}{nm} and \qty{528.8}{nm} SFG pulses can be clearly seen. When the \qty{1560}{nm} GDD and pulse chirp are negative ($b_{1}$ = \qty{-0.32}{THz/ps}) and thus opposite in direction to the \qty{800}{nm} pulse chirp, the change in instantaneous frequency of the sum-frequency pulse $b_{3}$ = \qty{0.13}{THz/ps}. For a pulse delay of \qty{0.5}{ps}, as shown by the plots of irradiance and frequency shift against absolute time after the pulses exit the nonlinear crystal, the instantaneous frequency of the sum-frequency pulse $\omega_{3}(t)$ is initially smaller than the centre frequency $\omega_{30}$, as shown by the initially negative frequency shift of the SFG pulse $\Delta \omega_{3}$. As the input pulses propagate through the crystal, $\Delta \omega_{3}$ increases but remains close to zero due to the slow change in instantaneous frequency. Thus, the input pulses satisfy the phase-matching condition of the nonlinear crystal for a prolonged period within the crystal, resulting in high sum-frequency conversion. However, when the pulse delay increases to \qty{5.0}{ps}, $\Delta \omega_{3}$ is instead initially greater than zero and continues to increase, resulting in poor phase-matching throughout the crystal and thus limited sum-frequency conversion as shown by the bottom irradiance plot.

In comparison, when the \qty{1560}{nm} GDD and pulse chirp are positive ($b_{1}$ = \qty{0.32}{THz/ps}), $b_{3}$ increases to \qty{0.77}{THz/ps} and thus $\omega_{3}(t)$ changes more rapidly. This is shown by the frequency shift plots at positive \qty{1560}{nm} GDD, where $\Delta \omega_{3}$ has a larger gradient in the outer wings of the SFG pulse. However, the region where $\Delta \omega_{3} \approx 0$ is wider than expected from Equation \ref{eq:instant-freq-sfg}. Additionally, the irradiance and frequency shift plots are very noisy, likely due to numerical errors in the simulation caused by the very narrow region where the phase-matching condition is satisfied and by the limits on the maximum possible resolution, especially due to the large number of time steps needed to resolve the \qty{800}{nm} pulse; further investigation is required to fully understand this behaviour. Due to the greater value of $b_{3}$ and the temporal walkoff of the pulses within the crystal as the \qty{1560}{nm} pulse catches up to the \qty{800}{nm} pulse, the input pulses will scan over the entire region where $\Delta \omega_{3} \approx 0$ for a large range of pulse delays. As such, significantly more sum-frequency radiation is generated at a pulse delay of \qty{5}{ps} compared to when the \qty{1560}{nm} chirp is negative. Additionally, at a pulse delay of \qty{0.5}{ps}, the narrow region where $\Delta \omega_{3} \approx 0$ reduces the length of time that the phase-matching condition is satisfied by the input pulses before they no longer overlap due to temporal walkoff, thus greatly limiting the maximum amount of sum-frequency radiation that can be generated.

\begin{figure}[t]
    \centering
    \includegraphics[width=0.8\textwidth]{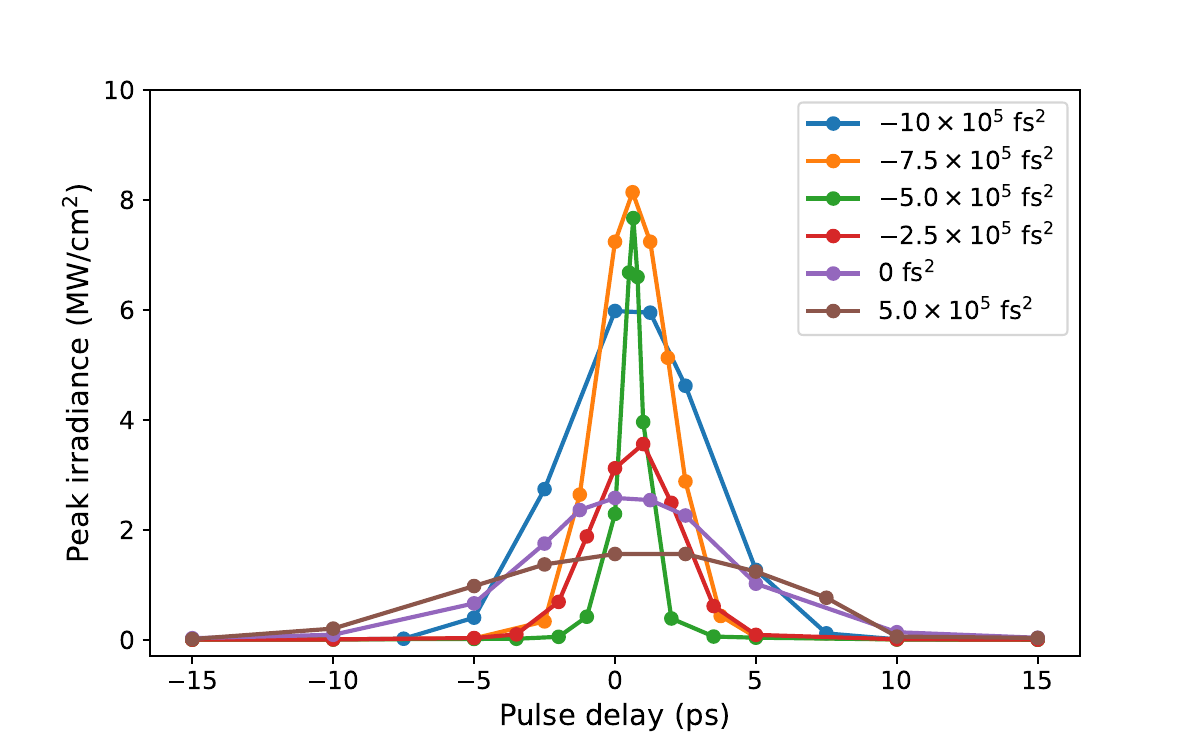}
    \caption{Plot of peak sum-frequency irradiance against pulse delay for different values of the \qty{1560}{nm} group delay dispersion (GDD). The GDD of the \qty{1560}{nm} pulse is changed from \qty{-10E5}{fs^{2}} to \qty{5.0E5}{fs^{2}}. Additional pulse parameters are given in Figure \ref{fig:snlo-parameters}.}
    \label{fig:peak-irradiance-pulse-delay-snlo}
\end{figure}

The effect of chirp on the peak sum-frequency irradiance can be further seen in Figure \ref{fig:peak-irradiance-pulse-delay-snlo}, where the peak sum-frequency irradiance for a given \qty{1560}{nm} pulse GDD is measured over a range of pulse delays between $\pm$\qty{15}{ps} to cross-correlate the \qty{1560}{nm} and \qty{800}{nm} pulses and generate a cross-correlation trace for various \qty{1560}{nm} pulse GDDs. The width of the cross-correlation trace decreases as the \qty{1560}{nm} pulse GDD becomes more negative due to the reduction in the chirp of the sum-frequency pulse. This leads to a smaller range of pulse delays where the phase-matching condition is satisfied, as discussed previously. In addition, the maximum peak irradiance also increases as the GDD becomes more negative. For negative \qty{1560}{nm} GDDs, the instantaneous frequency of the sum-frequency pulse $\omega_{3}(t)$ changes more slowly, resulting in better phase-matching at the pulse delay for which maximum sum-frequency generation occurs and thus greater peak sum-frequency irradiance. In comparison, for positive \qty{1560}{nm} GDDs, $\omega_{3}(t)$ changes more quickly and the phase-matching is not as strong, leading to reduced peak irradiance. As the GDD decreases from \qty{-5.0E5}{fs^{2}} to \qty{-7.5E5}{fs^{2}}, the maximum peak irradiance increases marginally but the cross-correlation trace width also increases. This is due to the \qty{1560}{nm} pulse intensity decreasing as the pulse duration increases, resulting in the outer wings of the pulse containing a greater proportion of the pulse energy. Once the frequency components satisfying the phase-matching condition are depleted, the sum-frequency process stops and the sum-frequency irradiance no longer grows, limiting the maximum sum-frequency irradiance that can be achieved. Additionally, due to the phase-matching region being wider as a result of the increased \qty{1560}{nm} pulse duration, depletion occurs over a large range of pulse delays, leading to increased cross-correlation trace width. As the GDD decreases further to \qty{-10E5}{fs^{2}}, the \qty{1560}{nm} pulse intensity continues to decrease, leading to quicker depletion of the \qty{1560}{nm} pulse, and the phase-matching region becomes even wider. This results in the peak sum-frequency irradiance decreasing and the cross-correlation trace width increasing. These simulations suggest that, due to the large positive GDD of the \qty{800}{nm} pulse, the \qty{1560}{nm} pulse requires a large negative GDD of around \qty{-5.0E5}{fs^{2}} to \qty{-7.5E5}{fs^{2}} to maximise the amplitude of the cross-correlation trace without the trace width increasing excessively.


\section{Experiment setup and results} \label{sec:experiment-setup}

\begin{figure}[t]
    \centering
    \includegraphics[width=0.65\textwidth]{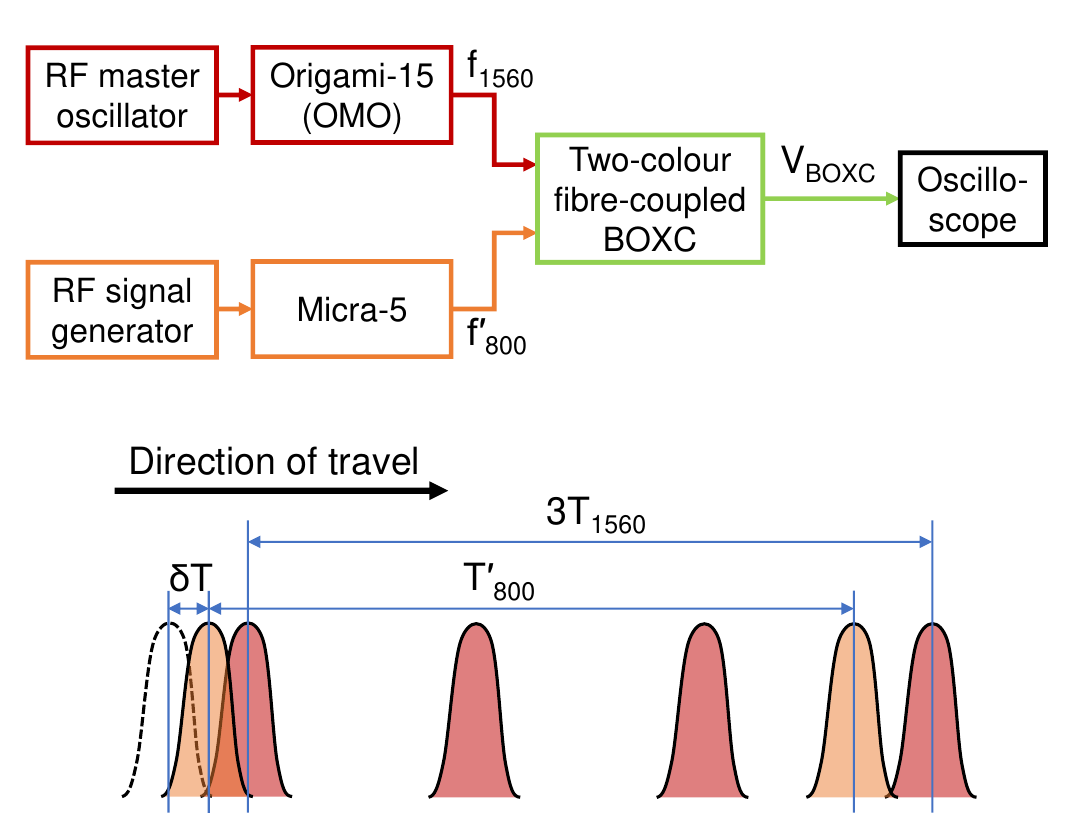}

    \caption{(a) Experiment setup for measuring the sensitivity of the two-colour fully fibre-coupled BOXC. (b) An illustration of the method used for generating the BOXC error signal. The difference in pulse periods ($\delta T << T_{800}'$) causes the temporal overlaps of the \qty{800}{nm} (orange) and \qty{1560}{nm} (red) pulse trains to change with every \qty{800}{nm} pulse, thus scanning over the entire width of the \qty{1560}{nm} pulse.}
    \label{fig:fibre-boxc-experiment-setup}
\end{figure}

Figure \ref{fig:fibre-boxc-experiment-setup} shows the setup to generate the BOXC voltage error curve and measure the sensitivity of the BOXC. The \qty{800}{nm} laser repetition rate is controlled by a signal generator that produces an RF signal at frequency $f_{800}' = f_{800} + \delta f = (f_{1560}/3) + \delta f$, where $\delta f << f_{800}, f_{1560}$. As mentioned in Section \ref{sec:principle-of-operation}, $f_{800}$ = \qty{83.292}{MHz} and $f_{1560}$ = \qty{249.875}{MHz}. This frequency offset causes the temporal overlaps of the \qty{800}{nm} and \qty{1560}{nm} pulse trains to slowly vary in time at a fixed rate. The change in arrival time of the \qty{800}{nm} pulse, $\delta T$, is given by

\begin{equation}
    \delta T = \frac{3}{f_{1560}} - \frac{1}{f_{800}'} = \frac{\delta f}{f_{800}(f_{800} + \delta{f})} \approx \frac{\delta f}{f_{800}^{2}} \; .
\end{equation}

\noindent The two PPLN waveguides in the BOXC then generate a train of evenly spaced cross-correlation traces, with each trace formed by multiple sum-frequency pulses as the \qty{800}{nm} and \qty{1560}{nm} pulse trains move through each other. Taking the voltage difference between the two cross-correlation traces generates the BOXC voltage error signal $V_{\text{BOXC}}$, which is recorded using an oscilloscope. The time recorded by the oscilloscope can then be converted to the pulse delay by the scan rate factor

\begin{equation}
    SR = \frac{\delta T}{T_{800}'} = \frac{\delta f}{f_{800}} \; ,
\end{equation}

\noindent which gives the amount that the pulse delay changes per unit of time elapsed. To use the full temporal range of the oscilloscope, but also to limit noise due to drifts in the repetition rate of the Micra-5, the frequency offset $\delta f$ is set to \qty{100}{Hz} for this experiment, giving a scan rate of \qty{1.2}{ps/\micro s}. Given that the bandwidth of the balanced photodetector output used in this experiment is \qty{4}{MHz} (see Section \ref{subsec:experiment-parameters}), the measurements of the BOXC error signal in Section \ref{subsec:two-colour-BOXC-sensitivity} will be averaged by a \qty{300}{fs}-long moving time window.


\subsection{Experiment parameters} \label{subsec:experiment-parameters}

Based on the results of the sum-frequency generation simulations in Section \ref{subsec:boxc-simulation}, an additional \qty{11}{m} of single-mode fibre was added to the \qty{1560}{nm} section before the EDFA input, leading to a total fibre length of \qty{35.1}{m}. Given that the \qty{1560}{nm} pulse duration before the waveguide entrance without additional fibre was measured to be \qty{1.1}{ps}, as mentioned in Section \ref{subsec:fibre-dispersion}, the additional \qty{11}{m} of fibre will cause the \qty{1560}{nm} pulse duration to increase to an expected value of \qty{3.9}{ps} with a GDD of \qty{-3.5E5}{fs^{2}}. Any additional fibre added to the system is \textit{in addition to} the \qty{11}{m} already added.

\begin{figure}[b!]
    \centering
    \subfigure[]{\includegraphics[width=0.495\textwidth]{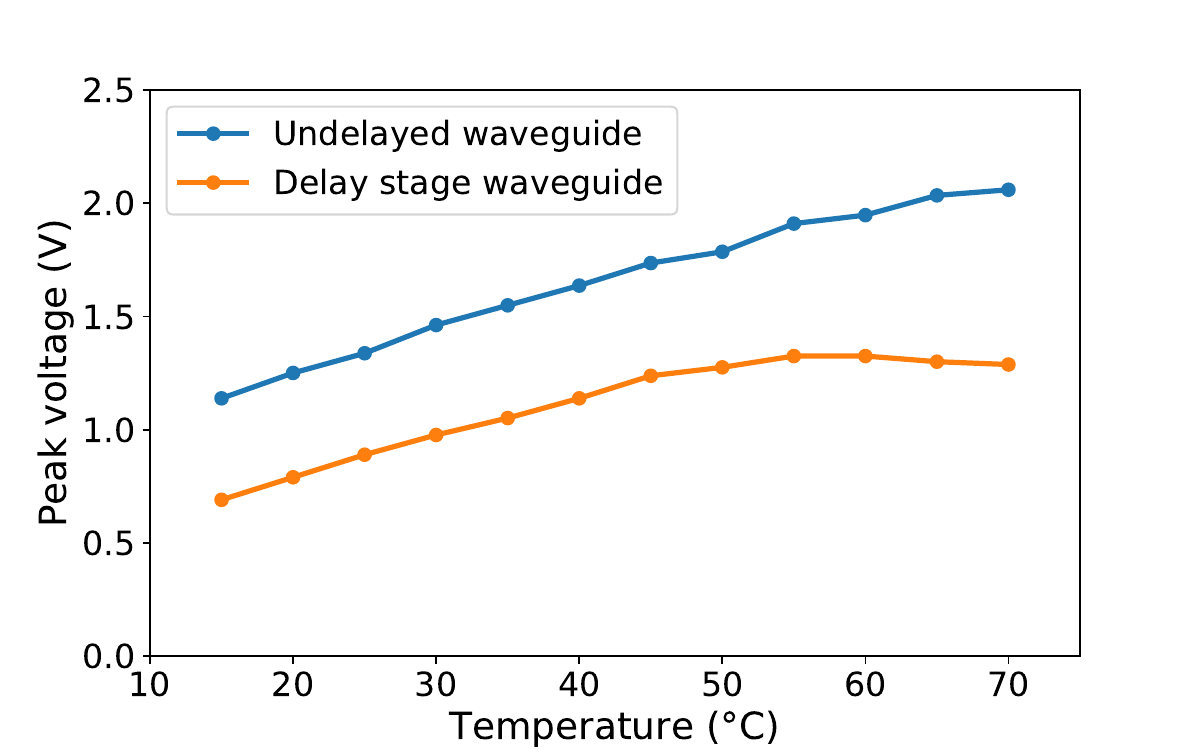}}
    \subfigure[]{\includegraphics[width=0.495\textwidth]{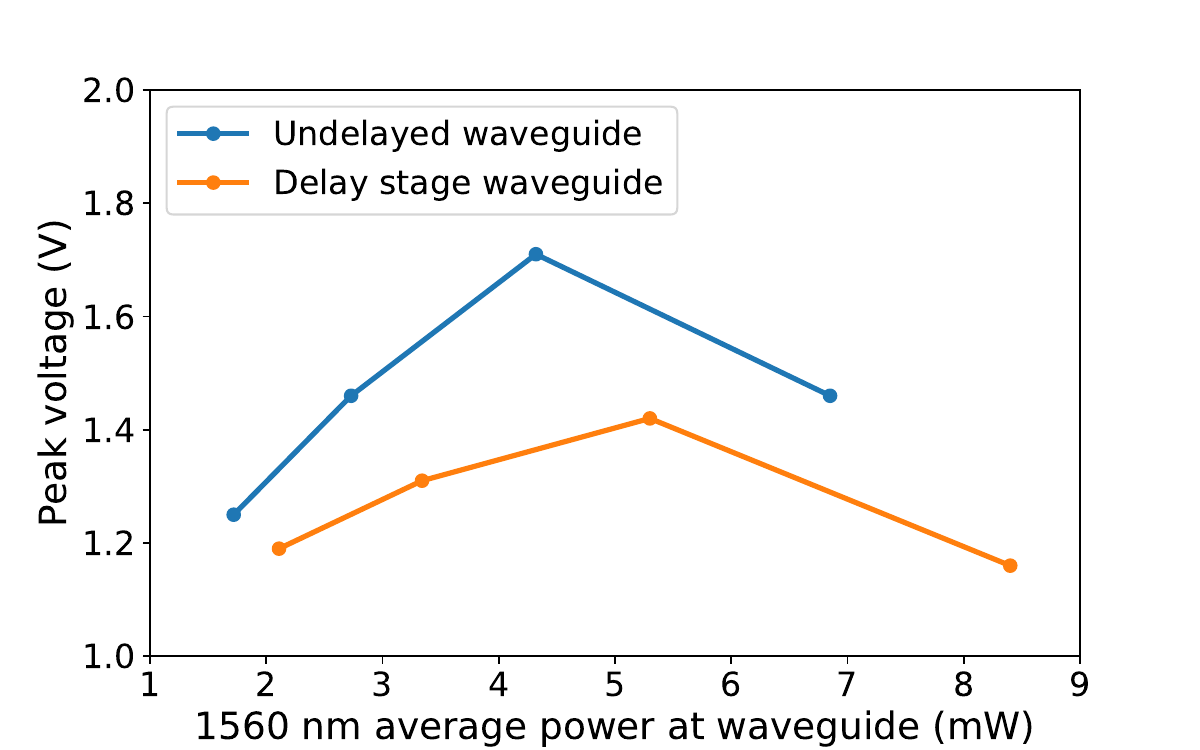}}
    \caption{(a) Plot of peak cross-correlation trace voltage against temperature for both PPLN waveguides. The maximum voltage occurs at a crystal temperature of \qty{70.0}{\degreeCelsius} for the undelayed waveguide and \qty{60.0}{\degreeCelsius} for the delay stage waveguide. (b) Plot of the peak cross-correlation trace voltage against \qty{1560}{nm} average power at the waveguide entrance for both PPLN waveguides. The temperatures of the undelayed waveguide and the delay stage waveguide were kept constant at \qty{70.0}{\degreeCelsius} and \qty{60.0}{\degreeCelsius} respectively. The \qty{800}{nm} average power at the waveguide entrance was also kept constant at \qty{3.5(0.1)}{mW} for the delay stage waveguide and \qty{1.8(0.1)}{mW} for the undelayed waveguide.}
    \label{fig:temp-scan}
\end{figure}

The balanced photodetector used in this experiment (Thorlabs PDB450A-AC) \cite{balanced-detector} consists of two well-matched photodiodes and a low-noise, high-speed transimpedance amplifier (TIA) with adjustable gain settings from \qty{E3}{V/A} to \qty{E7}{V/A}. The sum-frequency pulses from the waveguide that is not connected to the fibre delay stage (henceforth referred to as the `undelayed waveguide') and from the waveguide connected to the fibre delay stage (the `delay stage waveguide') are coupled into the `INPUT+' and `INPUT-' channels of the balanced photodetector respectively. The TIA then produces an output voltage proportional to the difference in photocurrent between the two input channels, which corresponds to the BOXC error signal in this experiment. The balanced photodetector also has two monitor outputs, allowing for the individual voltages from each input channel and hence the cross-correlation traces from each waveguide to be measured. For this experiment, the transimpedance (TI) gain for the BOXC error signal channel was set to \qty{E5}{V/A} to maximise the voltage whilst not saturating the detector. At this TI gain, the bandwidth of the balanced output is \qty{4}{MHz}. However, the TI gains of the input channels are not adjustable and are instead fixed at \qty{1.89E4}{V/A}. Thus, in the following results, the voltages of the cross-correlation traces are scaled to have the same TI gain of \qty{E5}{V/A} as the BOXC error signal to aid with visualisation of the data. The detector also has a responsivity of \qty{0.28}{A/W} at the centre wavelength of the sum-frequency radiation of \qty{528.8}{nm}.

Due to minute differences in manufacturing, the two waveguides have different optimal temperatures for maximum sum-frequency production of \qty{70.0}{\degreeCelsius} for the undelayed waveguide and \qty{60.0}{\degreeCelsius} for the delay stage waveguide. This is shown in Figure \ref{fig:temp-scan}(a), where the temperature of each waveguide was scanned from \qty{15.0}{\degreeCelsius} to \qty{70.0}{\degreeCelsius} at fixed \qty{800}{nm} average power and fixed \qty{1560}{nm} EDFA output power of \qty{10}{dBm}. Although it appears that the peak cross-correlation trace voltage from the undelayed waveguide would continue to increase at temperatures above \qty{70.0}{\degreeCelsius}, the waveguide controller prevents the TEC inside the waveguide housing from being set to temperatures above \qty{70.0}{\degreeCelsius}.

For the conversion efficiency and BOXC error signal measurements, the waveguides were set to the optimal temperatures found in Figure \ref{fig:temp-scan}(a) to maximise the cross-correlation trace voltage and the sensitivity of the BOXC. The \qty{1560}{nm} EDFA output power was varied from \qty{10}{dBm} to \qty{16}{dBm}, and the \qty{800}{nm} input power before the fibre port was set to \qty{16}{dBm}.


\subsection{Waveguide conversion efficiency} \label{subsec:conversion-efficiency}

As shown by Figure \ref{fig:temp-scan}(a), the conversion efficiency of the delay stage waveguide is less than that of the undelayed waveguide due to minute differences in manufacturing. This can be further seen in Figure \ref{fig:temp-scan}(b), where the \qty{1560}{nm} average power at the waveguide entrance was scanned between \qty{1}{mW} and \qty{9}{mW} by varying the EDFA output power from \qty{10}{dBm} to \qty{16}{dBm} and the peak cross-correlation trace voltage from each waveguide was measured. 

Due to the operating wavelengths of the 50:50 splitters being different to the laser wavelengths, the splitting ratios of both the \qty{1560}{nm} and \qty{800}{nm} splitters are not exactly 50:50, resulting in different average powers at both waveguide entrances. The \qty{800}{nm} average power at the entrance of the delay stage waveguide and the undelayed waveguide was measured to be \qty{3.5(0.1)}{mW} and \qty{1.8(0.1)}{mW} respectively, corresponding to respective power losses of \qty{10.6(0.1)}{dB} and \qty{13.4(0.1)}{dB} from before the fibre port to the waveguide entrance. From Figure \ref{fig:temp-scan}(b), the maximum cross-correlation trace voltages of \qty{1.42}{V} for the delay stage waveguide and \qty{1.71}{V} for the undelayed waveguide occurred when the output power of the \qty{1560}{nm} EDFA was set to \qty{14}{dBm}. At this EDFA output power, the measured \qty{1560}{nm} average power at the entrance of the delay stage waveguide and the undelayed waveguide was \qty{5.3(0.1)}{mW} and \qty{4.3(0.1)}{mW} respectively, corresponding to respective power losses of \qty{6.8(0.1)}{dB} and \qty{7.7(0.1)}{dB} from the EDFA output to the waveguide entrance. The predominant source of the power losses is the insertion loss of the various fibre-coupled components, especially because the components used were designed for different centre wavelengths of \qty{780}{nm} and \qty{1550}{nm}. 

Additionally, the peak voltages for both waveguides decreases as the \qty{1560}{nm} average power is increased further. This is likely due to self-phase modulation effects compressing the \qty{1560}{nm} pulse at higher EDFA output powers, thus reducing the magnitude of the negative GDD of the \qty{1560}{nm} pulse and leading to less intense sum-frequency pulses and reduced peak cross-correlation trace voltages as discussed in Section \ref{subsec:boxc-simulation}.

The conversion efficiency per unit length of the nonlinear crystal $\eta$ can be defined by

\begin{equation}
    \eta = \frac{P_{528}}{P_{1560} P_{800} l} \approx \frac{1}{l} \cdot \frac{0.88 P_{528}^{a}}{f_{528} \tau_{528}} \bigg(\frac{0.88 P_{1560}^{a}}{f_{1560} \tau_{1560}} \cdot \frac{0.88 P_{800}^{a}}{f_{800} \tau_{800}} \bigg)^{-1} \;,
    \label{eq:sfg-conversion-efficiency-per-cm}
\end{equation}

\noindent where $P$ and $P^{a}$ are the peak and average pulse powers respectively, $l$ = \qty{0.5}{cm} is the crystal length in centimetres, $f$ is the pulse repetition rate, and $\tau$ is the FWHM pulse duration \cite{Wilson_2011}. The factor of 0.88 is due to the sech$^{2}$ shape of the pulses \cite{Paschottasech2_shaped_pulses, 800nm-sech2-shape, 1560nm-sech2-shape}.

Table \ref{tab:conversion-efficiency-parameters} gives an overview of the laser parameters used to calculate the conversion efficiencies per unit length for both the delay stage waveguide and the undelayed waveguide. The \qty{528.8}{nm} average power exiting the waveguide is calculated using the expression

\begin{table}[htbp] 
\centering
\caption{\bf Laser Parameters for Conversion Efficiency Measurement}
\begin{tabular}{ccccc}
\hline
\begin{tabular}[c]{@{}c@{}}Wavelength \\ (nm)\end{tabular} & \begin{tabular}[c]{@{}c@{}}Avg. power, delay\\ stage waveguide\\ (mW)\end{tabular} & \begin{tabular}[c]{@{}c@{}}Avg. power, unde-\\ layed waveguide\\  (mW)\end{tabular} & \begin{tabular}[c]{@{}c@{}}Repetition rate\\ (MHz)\end{tabular} & \begin{tabular}[c]{@{}c@{}} FWHM pulse\\ duration\\ (ps)\end{tabular} \\ \hline
1560                                                       & $5.3 \pm 0.1$                                                                                  & $4.3 \pm 0.1$                                                                              & 249.985                                                           & $\sim 3.9^\textit{a}$                                                     \\
800.0                                                      &  $3.5 \pm 0.1$                                                                                  &  $1.8 \pm 0.1$                                                                                & 83.292                                                            & 39                                                              \\
528.8                                                      &             $0.057 \pm 0.001 ^\textit{b}$                                                                        &             $0.069 \pm 0.001 ^\textit{b}$                                                                        & $83.292$                                                            & $\sim 6^\textit{c}$                                                        \\ \hline
\end{tabular}
\label{tab:conversion-efficiency-parameters}
$^\textit{a}$Expected pulse duration calculated in Section \ref{subsec:experiment-parameters}. $^\textit{b}$Calculated from the peak cross-correlation trace voltage. $^\textit{c}$Estimated using SNLO.
\end{table}

\begin{equation}
    P_{528}^{a} = \frac{U}{G_{\text{TI}} R} \cdot 10^{\frac{\alpha}{10}}\; ,
\end{equation}

\noindent where $U$ is the peak cross-correlation trace voltage, $G_{\text{TI}}$ = \qty{1E5}{V/A} is the TI gain, $R$ = \qty{0.28}{A/W} is the detector responsivity at \qty{528.8}{nm}, and $\alpha$ = \qty{0.5(0.1)}{dB} is the coupling loss from the waveguide to the balanced photodetector due to the insertion losses of the WDM and the fibre mating sleeve. To estimate the pulse duration of the \qty{528.8}{nm} pulse exiting the waveguide, SNLO was used to simulate the nonlinear interaction at the experiment parameters in Table \ref{tab:conversion-efficiency-parameters} and the FWHM of the sum-frequency irradiance plot was taken. Using these values, the conversion efficiencies per unit length of the delay stage waveguide and the undelayed waveguide are found to be $\eta_{1}$ = \qty{4.4(0.3)}{\%\slash\watt\centi\metre} and $\eta_{2}$ = \qty{12.7(1.2)}{\%\slash\watt\centi\metre} respectively. 

Another measure of the conversion efficiency, $\eta'$, is given by

\begin{equation}
    \eta' = \frac{P_{528}}{\sqrt{(P_{800}P_{1560})}} \;,
    \label{eq:sfg-conversion-efficiency}
\end{equation}

\noindent where $P$ is the peak power \cite{1995Lublinski}. Using this equation and the parameters in Table \ref{tab:conversion-efficiency-parameters}, the conversion efficiencies of the delay stage waveguide and the undelayed waveguide are found to be $\eta_{1}'$ = \qty{4.7(0.2)}{\%} and $\eta_{2}'$ = \qty{8.8(0.4)}{\%} respectively. Despite the difference in conversion efficiency between the two crystals, the calculated values of $\eta'$ for both waveguides are more than 9 times greater than the $\sim$ \qty{0.5}{\%} conversion efficiency of BBO for this sum-frequency interaction \cite{schulz2015}. As a result, the fibre-coupled nonlinear crystal waveguides produce more intense sum-frequency radiation for the same laser power entering the nonlinear crystal, leading to potentially more sensitive BOXCs compared to what can be achieved using bulk nonlinear crystals.


\subsection{Two-colour fully fibre-coupled BOXC sensitivity} \label{subsec:two-colour-BOXC-sensitivity}

\begin{figure}[b!]
    \centering
    \includegraphics[width=0.75\textwidth]{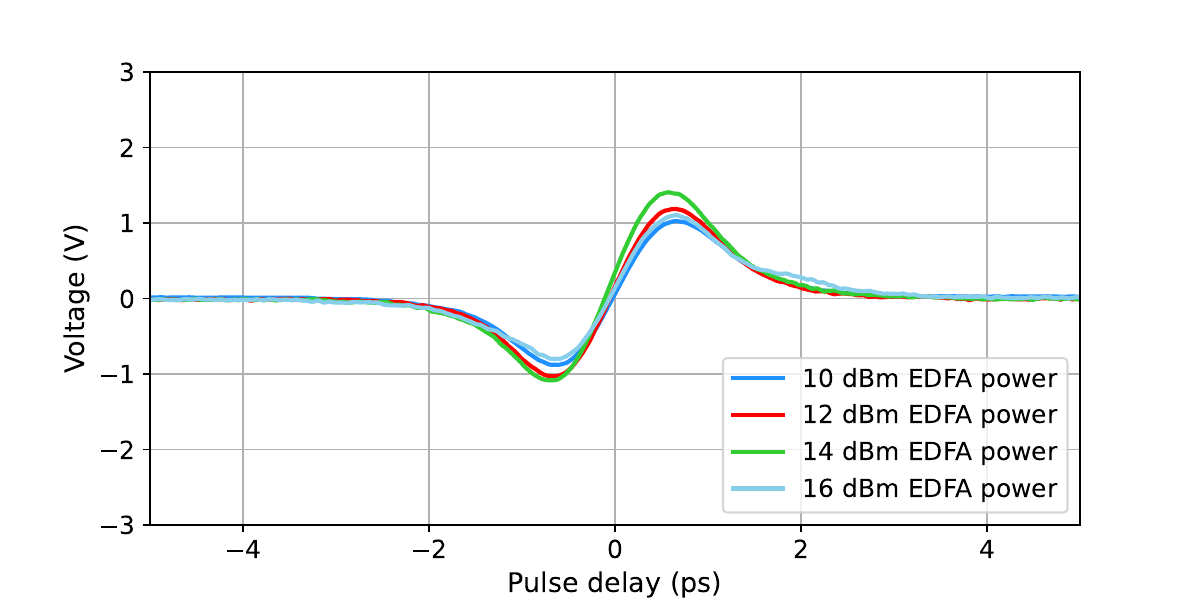}
    \caption{Effect of increasing EDFA output power on the BOXC voltage error signal. The \qty{800}{nm} power before the fibre port is \qty{16}{dBm}.}
    \label{fig:boxc-edfa-scan-no-dcf}
\end{figure}

Figure \ref{fig:boxc-edfa-scan-no-dcf} shows how the BOXC voltage error signal varies with EDFA power. The gradient of the error signal about the zero crossing point, which gives the sensitivity of the BOXC to pulse arrival time fluctuations, initially increases with increasing EDFA power, reaching a maximum sensitivity of \qty{3.1}{mV/fs} at an EDFA output power of \qty{14}{dBm}. However, at \qty{16}{dBm}, the effects of self-phase modulation reduce the magnitude of the GDD of the \qty{1560}{nm} pulses, resulting in less intense sum-frequency pulses and reduced BOXC sensitivity.

\begin{figure}[t]
    \centering
    \subfigure[]{\includegraphics[width=0.75\textwidth]{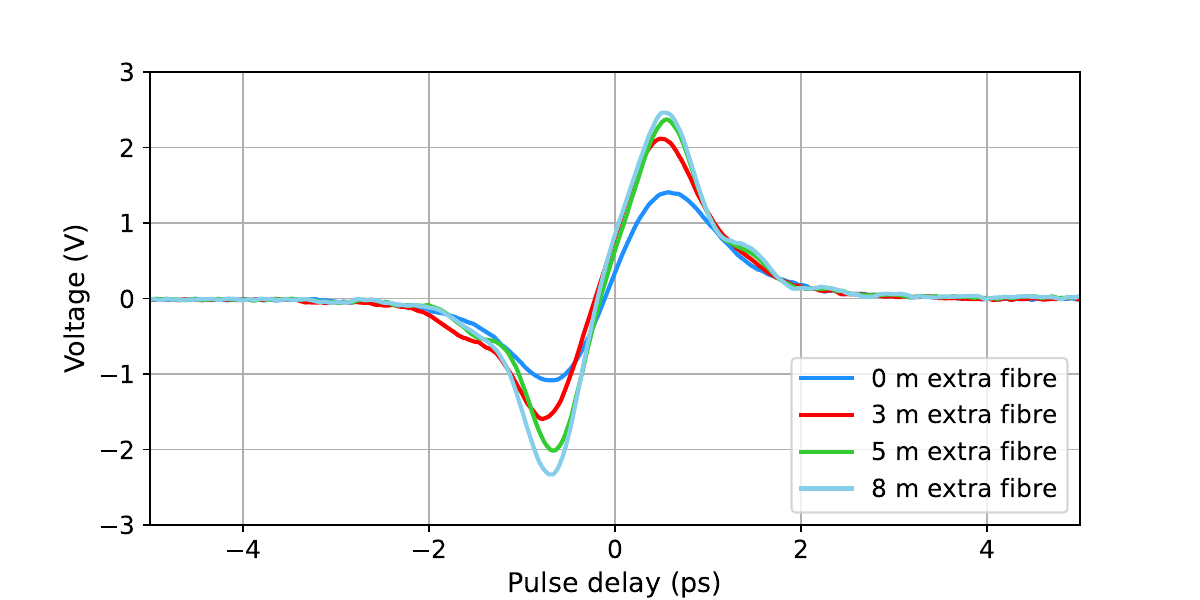}}
    \\
    \subfigure[]{\includegraphics[width=0.75\textwidth]{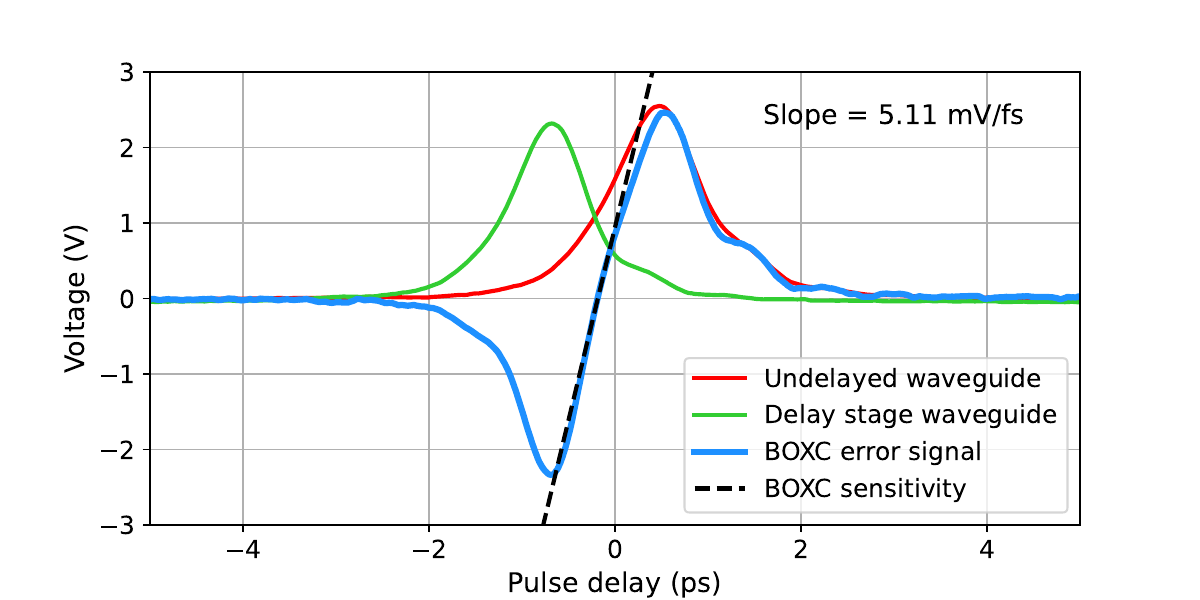}}
    \caption{(a) Effect of increasing \qty{1560}{nm} single-mode fibre length on the BOXC voltage error signal for an EDFA output power of \qty{14}{dBm}. (b) Cross-correlation traces and BOXC error signal for an EDFA output power of \qty{14}{dBm} and \qty{8}{m} of additional \qty{1560}{nm} single-mode fibre added before the EDFA input. The \qty{800}{nm} power before the fibre port is \qty{16}{dBm} for both figures.}
    \label{fig:14dBm-no-dcf-extra-fibre}
\end{figure}

By adding additional single-mode fibre to the \qty{1560}{nm} section of the BOXC, the GDD of the \qty{1560}{nm} pulse will become increasingly negative. This results in narrower and higher amplitude cross-correlation traces, as discussed in Section \ref{subsec:boxc-simulation}, both of which would increase the BOXC sensitivity. This can be seen in Figure \ref{fig:14dBm-no-dcf-extra-fibre}(a), where the EDFA output power is set to \qty{14}{dBm} and up to \qty{8}{m} of additional single-mode fibre is added before the EDFA. As additional fibre is added and the \qty{1560}{nm} GDD becomes increasingly negative, the amplitude of the cross-correlation traces increases and the FWHM of the traces decreases, as predicted by Figure \ref{fig:peak-irradiance-pulse-delay-snlo}. The maximum BOXC sensitivity occurs when \qty{8}{m} of additional fibre is added, decreasing the \qty{1560}{nm} GDD to \qty{-5.4E5}{fs^{2}}. This matches the results from the cross-correlation simulations in Section \ref{subsec:boxc-simulation}, where the cross-correlation trace is narrowest at a \qty{1560}{nm} GDD of \qty{-5.0E5}{fs^{2}}, although the simulations do not take nonlinear effects such as self-phase modulation into account.

The BOXC error signal at an EDFA output power of \qty{14}{dBm} with \qty{8}{m} of additional fibre added is shown in Figure \ref{fig:14dBm-no-dcf-extra-fibre}(b). For this measurement, the \qty{1560}{nm} and \qty{800}{nm} average powers entering the waveguides are the same as the values given in Table \ref{tab:conversion-efficiency-parameters}. The FWHM time widths of the cross-correlation traces are \qty{1.3}{ps} for the undelayed waveguide and \qty{1.1}{ps} for the delay stage waveguide. In comparison with the cross-correlation simulations in Section \ref{subsec:boxc-simulation}, the FWHM width of the \qty{-5.0E5}{fs^{2}} GDD cross-correlation trace is approximately \qty{0.8}{ps}. This difference is likely due to the effects of self-phase modulation reducing the GDD of the \qty{1560}{nm} pulses and causing the frequency shift of the \qty{1560}{nm} pulse to no longer be linear, whereas only pulses with initially linear frequency shifts can be modelled in the SNLO simulations performed in Section \ref{subsec:boxc-simulation}. The measured sensitivity of the two-colour fully fibre-coupled BOXC of \qty{5.11}{mV/fs} is 5 times higher than comparable bulk-optic two-colour BOXCs after accounting for TI gain and photodetector responsivity \cite{clli:ibic21-wepp05}, with potential for the sensitivity to increase further through compression of both the \qty{1560}{nm} and \qty{800}{nm} pulses.


\section{Conclusion} \label{sec:conclusion}

We have demonstrated, to the best of our knowledge, the first two-colour fully fibre-coupled balanced optical cross-correlator. This design is based upon sum-frequency generation between \qty{1560}{nm} and \qty{800}{nm} laser pulses using two PPLN waveguides. The conversion efficiencies of the two waveguides were measured to be $\eta_{1}'$ = \qty{4.7(0.2)}{\%} and $\eta_{2}'$ = \qty{8.8(0.4)}{\%}, significantly greater than bulk-optic BBO, although not accounting for optical losses in the two input fibres leading to the waveguide. Due to the large positive chirp of the \qty{800}{nm} pulse from dispersive broadening, it was found in both simulations and experiments that increasing the negative chirp of the \qty{1560}{nm} pulse by adding additional single-mode fibre increased the peak cross-correlation trace voltage and the sensitivity of the BOXC. Using these findings, a maximum BOXC sensitivity of \qty{5.11}{mV/fs} was achieved, 5 times greater than comparable bulk-optic BOXCs. Optimisations to the BOXC design will be made through reducing coupling losses by splicing the fibre components together, reducing insertion losses by using custom fibre components matched to the centre wavelengths of \qty{800}{nm} and \qty{1560}{nm}, compensating for fibre dispersion by reducing fibre lengths and by using dispersion-compensating fibre and other pulse compression methods, and reducing the overall footprint of the design. Future investigation of the performance of the two-colour fully fibre-coupled BOXC will involve measuring the environmental stability of the BOXC sensitivity, using the BOXC error signal to lock the \qty{800}{nm} laser to the \qty{1560}{nm} OMO, and measuring the resulting integrated timing jitter and long-term stability.


\begin{backmatter}

\bmsection{Funding}
This research was supported by the Science and Technology Facilities Council under grant ST/V001612/1 and by the studentship project 2489637.

\bmsection{Acknowledgement}
We thank the Femtosecond Lasers and Timing group at Daresbury Laboratory for the operation and maintenance of the Micra-5 and ORIGAMI-15 lasers and for their insights into the research. We also thank the research group at the University of Liverpool for their insights and expertise.

\bmsection{Disclosures}
The authors declare no conflicts of interest.

\bmsection{Data availability}
Data underlying the results presented in this paper may be obtained from the authors upon reasonable request.

\end{backmatter}


\bibliography{sources}

\end{document}